\documentclass[aps,floatfix,showpacs,twocolumn,superscriptaddress]{revtex4} 

\usepackage{graphicx} 
\usepackage{dcolumn} 
\usepackage{amsmath}

\newcommand{\sfrac}[2]{\mbox{\footnotesize $\displaystyle \frac{#1}{#2}$}} 

\begin{document} 
 
 
\title{%
Electromagnetic properties of ground and excited state pseudoscalar mesons} 
 
\author{A.\ H\"oll} 
\affiliation{Physics Division, Argonne National Laboratory, 
             Argonne, IL 60439-4843, U.S.A.} 
             
\author{A.\ Krassnigg}
\altaffiliation[Current Address:~]{Fachbereich Theoretische Physik, Universit\"at Graz, A-8010 Graz, Austria.}
\affiliation{Physics Division, Argonne National Laboratory, 
             Argonne, IL 60439-4843, U.S.A.} 

\author{P.\ Maris}
\affiliation{Department of Physics and Astronomy, University of Pittsburgh, Pittsburgh, PA 15260, U.S.A.} 
                                       
\author{C.D.\ Roberts} 
\affiliation{Physics Division, Argonne National Laboratory, 
             Argonne, IL 60439-4843, U.S.A.} 
\affiliation{Fachbereich Physik, Universit\"at Rostock, D-18051 Rostock, 
Germany} 

\author{S.\,V.\ Wright}
\affiliation{Physics Division, Argonne National Laboratory, 
             Argonne, IL 60439-4843, U.S.A.} 
             
\begin{abstract} 
\rule{0ex}{3ex} 
The axial-vector Ward-Takahashi identity places constraints on particular properties of every pseudoscalar meson.  For example, in the chiral limit all  pseudoscalar mesons, except the Goldstone mode, decouple from the axial-vector current.  Nevertheless, all neutral pseudoscalar mesons couple to two photons.  The strength of the $\pi_{n}^0 \gamma\gamma$ coupling, where $n=0$ denotes the Goldstone mode, is affected by the Abelian anomaly's continuum contribution.  The effect is material for $n\neq 0$.  The $\gamma^\ast \pi_n \gamma^\ast$ transition form factor, ${\cal T}_{\pi_n}(Q^2)$, is nonzero $\forall n$, and ${\cal T}_{\pi_n}(Q^2) \approx (4\pi^2/3) (f_{\pi_n}/Q^2)$ at large $Q^2$.  For all pseudoscalars but the Goldstone mode, this leading contribution vanishes in the chiral limit.  In this instance the ultraviolet power-law behaviour is $1/Q^4$ for $n\neq 0$, and we find numerically ${\cal T}_{\pi_1}(Q^2) \simeq (4\pi^2/3) (-\langle \bar q q \rangle/Q^4)$.  This subleading power-law behaviour is always present.  In general its coefficient is not simply related to $f_{\pi_n}$.  The properties of $n\neq 0$ pseudoscalar mesons are sensitive to the pointwise behaviour of the long-range piece of the interaction between light-quarks.  
\end{abstract} 
 
\maketitle 

\section{Introduction}
\label{sec:intro}
The known meson spectrum contains three pseudoscalars [$I^G (J^{PC}) = 1^- (0^{-+}) $], all with masses below $2\,$GeV \cite{pdg}: $\pi(140)$; $\pi(1300)$; and $\pi(1800)$.  The lightest of these, the pion [$\pi(140)$], is much studied and well understood as QCD's Goldstone mode.  It is the basic degree of freedom in chiral effective theories, and a veracious explanation of its properties requires an approach to possess a valid realisation of chiral symmetry and its dynamical breaking. 

The $\pi(1300)$ is broad, with a width of $200$ to $600\,$MeV.  In the framework of constituent-quark models it is usually interpreted as the pion's first radial excitation.  Namely, the $\pi(1300)$ is pictured as: an $I^G (J^{P}) L = 1^- (0^-)S$  $Q \bar Q$ meson, where $Q$ denotes a constituent-quark; and the first radially excited state of the $\pi(140)$ on a $Q \bar Q$ $n\, ^1\!S_0$ trajectory, where $n$ is the ``principal quantum number'' \cite{anisovich,norbury}.  

At first sight it might appear natural to interpret the $\pi(1800)$ as the third state on the $n\, ^1\!S_0$ trajectory.  However, in comparison with $\pi(1300)$, the $\pi(1800)$ is narrow, with a width of $207\pm 13\,$MeV, and has a decay pattern that may be consistent with its interpretation as a \emph{hybrid} meson in constituent-quark models \cite{page}.  This picture has the constituent-quarks' spins aligned to produce $S_{Q\bar Q} = 1$ with $J=0$ obtained by coupling $S_{Q\bar Q}$ to a spin-$1$ excitation of the confinement potential.

It is legitimate to ask for a unified theoretical understanding of these states and, indeed, the entire trajectory of pseudoscalar mesons.  This is a topical question; e.g., Refs.\,\cite{shakin,ji,barnes,yudichev,metsch2,bakker,krassnigg1,andreastunl,%
lc03,andreasrapid}, and it is easy to identify at least one reason why.  In the context of a constituent-quark model Hamiltonian a subset, if not all, of the pseudoscalar mesons form a $Q\bar Q$ $n\, ^1\!S_0$ trajectory.  In this framework the support possessed at long-range by the bound state's wave function grows with increasing $n$.  Hence the properties of radially excited states become increasingly sensitive to the manner by which confinement is expressed in the potential.  As we have already noted, in this same context a definition and representation of hybrid mesons requires that explicit excitation of the confinement potential be included as an additional degree of freedom.  Seen from this perspective one may anticipate that the properties of all the heavier pseudoscalar mesons are likely to be sensitive to the long-range part of the interaction between light-quarks in QCD, whether they be radial excitations or hybrid mesons.  This suggests that the study of their properties can provide a map of what might be called the confinement potential between light-quarks.  (NB. The information obtained thereby is complementary to that gathered in studies of axial-vector mesons \cite{ericaxial,a1b1,burdenpichowsky,jarecke,peteraxial}, which in constituent-quark models are interpreted as orbital excitations of the $\pi$- and $\rho$-mesons.)

It is not possible to accurately describe pseudoscalar mesons using a framework that fails to respect the axial-vector Ward-Takahashi identity.  For example, chiral symmetry and its dynamical breakdown force the leptonic decay constant of every pseudoscalar meson, \emph{except} the Goldstone mode, to vanish in the chiral limit \cite{yudichev,krassnigg1,andreastunl,lc03,andreasrapid}.  Herein we therefore employ QCD's Dyson-Schwinger equations (DSEs) (modern applications are reviewed in Refs.\,\cite{bastirev,reinhardrev,pieterrev}) for which a systematic, Poincar\'e covariant and symmetry preserving treatment of quark-antiquark bound states has been established \cite{bender,detmold,mandarvertex}.  To provide exemplars we will focus primarily on the $\pi(140)$ and the next-lightest pseudoscalar state.  Nonetheless,  the exact results will apply to all elements on the pseudoscalar meson trajectory.  

It is noteworthy that in Poincar\'e covariant quantum field theory all bound states with given quantum numbers; e.g., ($I^G$, $J^{PC}$), are described by the same homogeneous Bethe-Salpeter equation (BSE).  This is kindred to the statement that all interpolating fields with the same quantum numbers are on-shell equivalent, a fact which is apparent in numerical simulations of lattice-QCD; e.g., Ref.\,\cite{hedditch}.  Hence a given homogeneous BSE yields the mass and Bethe-Salpeter amplitude of every bound state in the channel specified by ($I^G$, $J^{PC}$).  

In a confining theory a given $J^P$ trajectory will likely contain a countable infinity of bound states.  The lowest mass member of the trajectory is conventionally described as the ground state.  All other members may reasonably be described as excited states.  The radial excitation of a state with a given $J^P$ preserves this total-momentum\,$+\,$parity assignment.  However, it may be distinguished from the ground state by the pointwise behaviour of its Bethe-Salpeter amplitude, which when analysed appropriately exhibits a finite number of zeros.  As in quantum mechanics, the number of zeros can be associated with a principal quantum number $n$.  Studies of pseudoscalar mesons show that the ground state amplitude has no zeros and can therefore be associated with $n=0$.  The amplitude of the next highest mass pseudoscalar possesses one zero and is therefore identified with $n=1$; e.g., \cite{yudichev,krassnigg1,andreastunl,lc03,andreasrapid}.  In simple models, this pattern continues \cite{metsch2,bakker}.  

It may be that hybrid mesons, if they exist, can likewise be identified through the pointwise behaviour of their Bethe-Salpeter amplitudes.  For example, a solution of the pseudoscalar BSE, heavier than the first radial excitation, whose Bethe-Salpeter amplitude exhibits both: a pattern of zeros which does not match that associated with radial excitations; and relationships between component functions in the Bethe-Salpeter amplitude different from those present in the lower mass solutions, would appear a reasonable hybrid candidate.  

Of course, Bethe-Salpeter amplitudes are not themselves observable and the experimental categorisation of ground, and excited and putative hybrid states proceeds via analysis of their decay patterns.  Notwithstanding this, the order in those decay patterns is determined in large part by the Bethe-Salpeter amplitudes' pointwise behaviour.  We therefore anticipate that a natural distinction between straightforward radial excitations and hybrids may be possible without recourse to a constituent-quark model basis.  

In Sec.\,\ref{gapbse} we recapitulate on aspects of the DSEs and truncation scheme that are relevant to our study.  The Abelian anomaly features in Sec.\,\ref{exact}, wherein exact results are derived regarding the coupling of pseudoscalar mesons to two photons.  We outline a renormalisation-group-improved model of the quark-antiquark scattering kernel in Sec.\,\ref{model}.  It is used in that section to illustrate the exact results and explore effects of the model's realisation of light-quark confinement on, e.g., bound state charge radii.  Section~\ref{epilogue} is an epilogue.

\section{BETHE-SALPETER AND GAP EQUATIONS}
\label{gapbse}
A Poincar\'e covariant and symmetry preserving treatment of quark-antiquark bound states can be based on the homogeneous Bethe-Salpeter equation (BSE) \cite{fn:Eucl}
\begin{equation}
\label{bse1}
[\Gamma(k;P)]_{tu} = \int^\Lambda_q [\chi(q;P)]_{sr}\, K_{rs}^{tu}(q,k;P)\,,
\end{equation}
where: $k$ is the relative and $P$ the total momentum of the constituents; $r$,\ldots,\,$u$ represent colour, Dirac and flavour indices; 
\begin{equation}
\label{definechi}
\chi(q;P)= S(q_+) \Gamma(q;P) S(q_-)\,,
\end{equation}
$q_\pm = q\pm P/2$; and $\int^\Lambda_q$ represents a Poincar\'e invariant regularisation of the integral, with $\Lambda$ the regularisation mass-scale \cite{mrt98,mr97}.  (We shall subsequently describe regularisation explicitly.)  In Eq.\,(\ref{bse1}), $S$ is the renormalised dressed-quark propagator and $K$ is the fully amputated dressed-quark-antiquark scattering kernel; namely, it is the sum of all diagrams that cannot be disconnected by cutting two fermion lines.  The product $(SS) K$ is a renormalisation point invariant.  Hence, when the kernel is expressed completely in terms of renormalised Schwinger functions, the homogeneous BSE's solution is independent of the regularisation mass-scale, which may be removed; viz., $\Lambda \to \infty$.

In a given channel the homogeneous BSE only has solutions for particular, separated values of $P^2$: $P^2=-m_n^2$, where $m_n$ is a bound state's mass,  whereat $\Gamma_n(k;P)$ is that bound state's Bethe-Salpeter amplitude.  In the flavour nonsinglet pseudoscalar channel the lowest mass solution is associated with the $\pi(140)$.  It is precisely QCD's Goldstone mode \cite{mrt98}, and we denote it by a value of $n=0$.  The homogeneous BSE next possesses a $J^{PC}=0^{-+}$ solution when $P^2$ assumes a value associated with the mass of the $\pi(1300)$.  We label this state by $n=1$.  In the study of this meson in Ref.\,\cite{andreastunl} the Tchebychev moments of the Lorentz scalar functions that appear in the matrix-valued Bethe-Salpeter amplitude each exhibit a single zero.  It can therefore be described as a radially excited state.  (NB. Hereafter the subscript $n$ is merely a counter labelling states of increasing mass: $m_0<m_1<m_2<\ldots$, etc.)  

The pattern of isolated solutions continues so that in principle one may obtain the mass and amplitude of every pseudoscalar meson from Eq.\,(\ref{bse1}).  Herein we will exploit this in comparing properties of the two lowest-mass flavour-nonsinglet $J^{PC}=0^{-+}$ mesons just described.  

The dressed-quark propagator appearing in the BSE's kernel is determined by the renormalised gap equation 
\begin{eqnarray}
S(p)^{-1} & =&  Z_2 \,(i\gamma\cdot p + m^{\rm bm}) + \Sigma(p)\,, \label{gendse} \\
\Sigma(p) & = & Z_1 \int^\Lambda_q\! g^2 D_{\mu\nu}(p-q) \frac{\lambda^a}{2}\gamma_\mu S(q) \Gamma^a_\nu(q,p) , \label{gensigma}
\end{eqnarray}
wherein: $D_{\mu\nu}$ is the dressed-gluon propagator, $\Gamma_\nu(q,p)$ is the dressed-quark-gluon vertex, and $m^{\rm bm}$ is the $\Lambda$-dependent current-quark bare mass.  The quark-gluon-vertex and quark wave function renormalisation constants, $Z_{1,2}(\zeta^2,\Lambda^2)$, depend on the gauge parameter, the renormalisation point, $\zeta$, and the regularisation mass-scale.  A Poincar\'e invariant regularisation of the integral is essential and, since pseudoscalar mesons are our focus, we employ a Pauli-Villars scheme.  That is implemented in Eq.\,(\ref{gendse}) by considering the quarks as minimally anticoupled ($g^{PV}=ig$) to additional massive gluons ($m_g^{PV}=\Lambda$).  This effects a tempering of the integrand, which is expressed via a modification of the gluon propagator's ultraviolet behaviour:
\begin{equation}
\frac{1}{(p-q)^2} \to \frac{1}{(p-q)^2} - \frac{1}{(p-q)^2+\Lambda^2}\,,
\end{equation}
and regulates the integral's superficial linear divergence.  

The gap equation's solution has the form 
\begin{eqnarray} 
 S(p)^{-1} & = & i \gamma\cdot p \, A(p^2,\zeta^2) + B(p^2,\zeta^2) \,,\\ 
& =& \frac{1}{Z(p^2,\zeta^2)}\left[ i\gamma\cdot p + M(p^2)\right] . 
\label{sinvp} 
\end{eqnarray} 
It is obtained from Eq.\,(\ref{gendse}) augmented by the renormalisation condition
\begin{equation}
\label{renormS} \left.S(p)^{-1}\right|_{p^2=\zeta^2} = i\gamma\cdot p +
m(\zeta)\,,
\end{equation}
where $m(\zeta)$ is the renormalised (running) current-quark mass: 
\begin{equation}
Z_2(\zeta^2,\Lambda^2) \, m^{\rm bm}(\Lambda) = Z_4(\zeta^2,\Lambda^2) \, m(\zeta)\,,
\end{equation}
with $Z_4$ the Lagrangian mass renormalisation constant.  At one-loop order in perturbative QCD 
\begin{equation}
m(\zeta) = \frac{\hat m}{(\ln \zeta/\Lambda_{\rm QCD})^{\gamma_m}}\,,
\end{equation}
with $\gamma_m= 12/(33-2 N_f)$, where $N_f$ is the number of active current-quark flavours, and $\hat m$ is the renormalisation-point-invariant current-quark mass.  The chiral limit is unambiguously defined by setting $\hat m = 0$ \cite{mrt98,mr97,langfeld}, which is equivalent to the requirement
\begin{equation}
\label{limchiral}
Z_2(\zeta^2,\Lambda^2) \, m^{\rm bm}(\Lambda) \equiv 0 \,,\; \forall \Lambda \gg \zeta \,.
\end{equation}

The behaviour and features of the solution of QCD's gap equation are reviewed in Refs.\,\cite{bastirev,reinhardrev,pieterrev}.  It is a longstanding prediction of DSE studies that the dressed-quark propagator is strongly dressed at infrared length-scales, namely, $p^2\lesssim 2\,$GeV$^2$ and that this is materially important in explaining a wide range of hadron properties \cite{pieterrev}.  Indeed, an enhancement of the mass function, $M(p^2)$, is central to the appearance of a constituent-quark mass-scale and an existential prerequisite for Goldstone modes.  The DSE results have been confirmed in numerical simulations of lattice-regularised QCD \cite{bowman} and the conditions have been explored under which pointwise agreement between DSE results and lattice simulations may be obtained \cite{bhagwat,maris,bhagwat2}.

The $I^G (J^{PC}) = 1^- (0^{-+})$ trajectory contains the pion, whose properties are fundamentally governed by the phenomenon of dynamical chiral symmetry breaking (DCSB).  One expression of the chiral properties of QCD is the axial-vector Ward-Takahashi identity 
\begin{eqnarray}
\nonumber
P_\mu \Gamma_{5\mu}^j(k;P) & =& S^{-1}(k_+) i \gamma_5\frac{\tau^j}{2}
+  i \gamma_5\frac{\tau^j}{2} S^{-1}(k_-)\\
&& - \, 2i\,m(\zeta) \,\Gamma_5^j(k;P) ,
\label{avwtim}
\end{eqnarray}
which we have here written for two quark flavours, each with the same current-quark mass: $\{\tau^i:i=1,2,3\}$ are flavour Pauli matrices.  In Eq.\,(\ref{avwtim}), $\Gamma_{5\mu}^j(k;P)$ is the axial-vector vertex:  
\begin{eqnarray}
\nonumber
\left[\Gamma^j_{5\mu}(k;P)\right]_{tu}
 & = &  Z_2 \left[\gamma_5\gamma_\mu \frac{\tau^j}{2} \right]_{tu}\\
 &+& \int^\Lambda_q
[\chi^j_{5\mu}(q;P)]_{sr} K_{tu}^{rs}(q,k;P)\,,
\label{avbse}
\end{eqnarray}
and $\Gamma_5^j(k;P)$ is the pseudoscalar vertex
\begin{eqnarray}
\nonumber
\left[\Gamma_{5}(k;P)\right]_{tu}
 & = &  Z_4 \left[\gamma_5 \frac{\tau^j}{2}\right]_{tu}\\
 &+&  \int^\Lambda_q
[\chi^j_{5}(q;P)]_{sr} K_{tu}^{rs}(q,k;P)\,.
\label{psbse}
\end{eqnarray}

The quark propagator, axial-vector and pseudoscalar vertices are all expressed via integral equations; i.e., DSEs.  Equation~(\ref{avwtim}) is an exact statement about chiral symmetry and the pattern by which it is broken.  Hence it must always be satisfied.  Since that cannot credibly be achieved through fine tuning, the distinct kernels of Eqs.\,(\ref{gendse}), (\ref{gensigma}), (\ref{avbse}), (\ref{psbse}) must be intimately related.  Any theoretical tool employed in calculating properties of the pseudoscalar and pseudovector channels must preserve that relationship if the results are to be both quantitatively and qualitatively reliable.

While a weak coupling expansion of the DSEs yields perturbation theory and satisfies this constraint, that truncation scheme is not useful in the study of bound states nor of other intrinsically nonperturbative phenomena; such as confinement and DCSB.  Fortunately at least one nonperturbative, systematic and symmetry preserving scheme exists. (References\,\cite{detmold,mandarvertex} give details.)  This entails that the full implications of Eq.\,(\ref{avwtim}) can be elucidated and illustrated.

Unless there is a reason for the residue to vanish, every isovector pseudoscalar meson appears as a pole contribution to the axial-vector and pseudoscalar vertices \cite{mrt98}: 
\begin{eqnarray}
\nonumber
\left. \Gamma_{5 \mu}^j(k;P)\right|_{P^2+m_{\pi_n}^2 \approx 0}&=&   \frac{f_{\pi_n} \, P_\mu}{P^2 + 
m_{\pi_n}^2} \Gamma_{\pi_n}^j(k;P) \\
& & + \; \Gamma_{5 \mu}^{j\,{\rm reg}}(k;P) \,, \label{genavv} \\
\nonumber 
\left. i\Gamma_{5 }^j(k;P)\right|_{P^2+m_{\pi_n}^2 \approx 0}
&=&   \frac{\rho_{\pi_n}(\zeta) }{P^2 + 
m_{\pi_n}^2} \Gamma_{\pi_n}^j(k;P)\\
& & + \; i\Gamma_{5 }^{j\,{\rm reg}}(k;P) \,; \label{genpv} 
\end{eqnarray}
viz., each vertex may be expressed as a simple pole plus terms regular in the neighbourhood of this pole, with $\Gamma_{\pi_n}^j(k;P)$ representing the bound state's canonically normalised Bethe-Salpeter amplitude: 
\begin{eqnarray} 
\nonumber
\lefteqn{
\Gamma_{\pi_n}^j(k;P) = \tau^j \gamma_5 \left[ i E_{\pi_n}(k;P) + \gamma\cdot P F_{\pi_n}(k;P) \right. }\\
&+& \left.
    \gamma\cdot k \,k \cdot P\, G_{\pi_n}(k;P) + 
\sigma_{\mu\nu}\,k_\mu P_\nu \,H_{\pi_n}(k;P)  \right] \! ; \label{genpibsa} \end{eqnarray}
and
\begin{eqnarray} 
\label{fpin} f_{\pi_n} \,\delta^{ij} \,  P_\mu &=& Z_2\,{\rm tr} \int^\Lambda_q 
\sfrac{1}{2} \tau^i \gamma_5\gamma_\mu\, \chi^j_{\pi_n}(q;P) \,, \\
\label{cpres} i  \rho_{\pi_n}\!(\zeta)\, \delta^{ij}  &=& Z_4\,{\rm tr} 
\int^\Lambda_q \sfrac{1}{2} \tau^i \gamma_5 \, \chi^j_{\pi_n}(q;P)\,.
\end{eqnarray} 
The residues expressed in Eqs.\,(\ref{fpin}) and (\ref{cpres}), are gauge invariant and cutoff independent.  

For an elementary pseudoscalar meson, $F_{\pi_n}(k;P)\equiv 0 \equiv G_{\pi_n}(k;P) \equiv H_{\pi_n}(k;P)$ in Eq.\,(\ref{genpibsa}).  The first two of these functions can be described as characterising the pseudoscalar meson's pseudovector components; and the last, its pseudotensor component.  The associated Dirac structures necessarily occur in a Poincar\'e covariant bound state description: they signal the presence of quark orbital angular momentum.

Equation\,(\ref{avwtim}) combined with Eqs.\,(\ref{genavv}) -- (\ref{cpres}) yields \cite{mrt98,mr97}
\begin{equation}
\label{gmorgen} f_{\pi_n} m_{\pi_n}^2 = 2 \, m(\zeta)  \, 
\rho_{\pi_n}(\zeta)\,; 
\end{equation}
i.e., an identity valid: for every flavour nonsinglet $0^-$ meson; and irrespective of the magnitude of the current-quark mass \cite{mishasvy}.  In the chiral limit additional information about the ground state pseudoscalar ($n=0$) is available; namely, an array of quark-level Goldberger-Treiman relations \cite{mrt98}
\begin{eqnarray}
\label{bwti} f_{\pi_0}^0 E_{\pi_0}(k;0)  &= &  B(k^2)\,, \\
 F_R(k;0) +  2 \, f_{\pi_0}^0 F_{\pi_0}(k;0)  & = & A(k^2)\,,\label{fwti}\\
G_R(k;0) +  2 \,f_{\pi_0}^0 G_{\pi_0}(k;0)    & = & 2 A^\prime(k^2)\,,\label{gwti}\\
H_R(k;0) +  2 \,f_{\pi_0}^0 H_{\pi_0}(k;0)    & = & 0\,, \label{hwti}
\end{eqnarray}
where $F_R$, $G_R$, $H_R$ are, respectively, the coefficient functions of $\gamma_5 \gamma_\mu$, $\gamma\cdot k k_\mu$, $\sigma_{\mu\nu} k_\nu$ in $\Gamma_{5 \mu}^{j\,{\rm reg}}(k;P)$ and 
\begin{equation}
f_{\pi_n}^0 := \lim_{\hat m \to 0}\, f_{\pi_n} .
\end{equation}
Equations~(\ref{bwti}) -- (\ref{hwti}) are a pointwise consequence of DCSB and a pointwise expression of Goldstone's theorem.  They can be used to show  
\begin{equation}
\rho_{\pi_0}^0(\zeta) := \lim_{\hat m \to 0}\,\rho(\zeta) = -\frac{1}{f^0_{\pi_0} } \langle \bar q q \rangle^0_\zeta\,,
\end{equation}
wherein  
\begin{equation} 
\label{qbq0} \,-\,\langle \bar q q \rangle_\zeta^0 = \lim_{\Lambda\to \infty} 
Z_4(\zeta^2,\Lambda^2)\, N_c \, {\rm tr}_{\rm D}\int^\Lambda_q\!
S^{0}(q,\zeta)\,,  
\end{equation} 
is the vacuum quark condensate \cite{langfeld}.  It is now plain from Eq.\,(\ref{gmorgen}) that in the neighbourhood of $\hat m = 0$
\begin{equation}
(f_{\pi_0}^0)^2 m_{\pi_0}^2 = -\, 2 \, m(\zeta)  \, \langle \bar q q \rangle_\zeta^0\,;
\end{equation} 
viz., the Gell-Mann--Oakes--Renner relation is a corollary of Eq.\,(\ref{gmorgen}).

\section{Two photon coupling of Pseudoscalar Mesons: Exact Results} 
\label{exact}
\subsection{Abelian anomaly}
To be concrete we will begin by considering the two-photon coupling as expressed via the renormalised triangle diagrams:
\begin{eqnarray}
\nonumber T^3_{5\mu\nu\rho}(k_1,k_2) &=& {\rm tr}\int_\ell^M {\cal S}(\ell_{0+}) \, \Gamma^3_{5\rho}(\ell_{0+},\ell_{-0}) \, {\cal S}(\ell_{-0}) \\%
\nonumber & \times&  \, i{\cal Q}\Gamma_\mu(\ell_{-0},\ell) \, {\cal S}(\ell) \, i {\cal Q}\Gamma_\nu(\ell,\ell_{0+})\,,\\
&& \label{Tmnr}\\
\nonumber T^3_{5\mu\nu}(k_1,k_2) &=& {\rm tr}\int_\ell^M {\cal S}(\ell_{0+}) \, \Gamma^3_{5}(\ell_{0+},\ell_{-0}) \, {\cal S}(\ell_{-0}) \\
\nonumber &\times&  \, i{\cal Q}\Gamma_\mu(\ell_{-0},\ell) \, {\cal S}(\ell) \, i {\cal Q}\Gamma_\nu(\ell,\ell_{0+})\,,\\
&& \label{Pmnr}
\end{eqnarray}
where $\ell_{\alpha\beta}=\ell+\alpha k_1+\beta k_2$, the electric charge matrix ${\cal Q}={\rm diag}[e_u,e_d]=e\,{\rm diag}[2/3,-1/3]$, ${\cal S}= {\rm diag}[S_u,S_d]$ and 
\begin{equation}
\left[\Gamma_{\mu}(k;P)\right]_{tu} = Z_2 \left[\gamma_\mu  \right]_{tu}\\
+ \int^\Lambda_q
[\chi^j_{\mu}(q;P)]_{sr} K_{tu}^{rs}(q,k;P) 
\end{equation}
is the renormalised dressed-quark-photon vertex.  

The bare axial-vector--vector--vector vertex exhibits a superficial linear divergence and, as with all other Schwinger functions, it must be rigorously defined via a Poincar\'e invariant regularisation scheme.  In this case an appropriate Pauli-Villars prescription corresponds to minimally anticoupling the photon to additional flavoured quarks with a large mass $m^{PV}=M$.  To elucidate, we introduce
\begin{eqnarray}
\nonumber \lefteqn{
\tilde T^3_{5\mu\nu\rho}(k_1,k_2;\hat m) := {\rm tr}\int_\ell {\cal S}_{\hat m}(\ell_{0+}) \, \Gamma^{3\,\hat m}_{5\rho}(\ell_{0+},\ell_{-0}) }\\
\nonumber
&& \times \, {\cal S}_{\hat m}(\ell_{-0})\, i {\cal Q} \Gamma^{\hat m}_\mu(\ell_{-0},\ell) \, {\cal S}_{\hat m}(\ell) \, i {\cal Q}\Gamma^{\hat m}_\nu(\ell,\ell_{0+})\,,\\ \label{PVregd}
\end{eqnarray}
wherein the current-quark-mass dependence is explicit, so that Eq.\,(\ref{Tmnr}) can rigorously be written as
\begin{equation}
T^3_{5\mu\nu\rho}(k_1,k_2;\hat m) = \tilde T^3_{5\mu\nu\rho}(k_1,k_2;\hat m) - \tilde T^3_{5\mu\nu\rho}(k_1,k_2;M)\,,
\end{equation}
with $M \to \infty$ as the last step in the calculation. 

\begin{widetext}
\begin{figure}[h]
\begin{center}
\hspace*{0em}\includegraphics[width=0.99\textwidth]{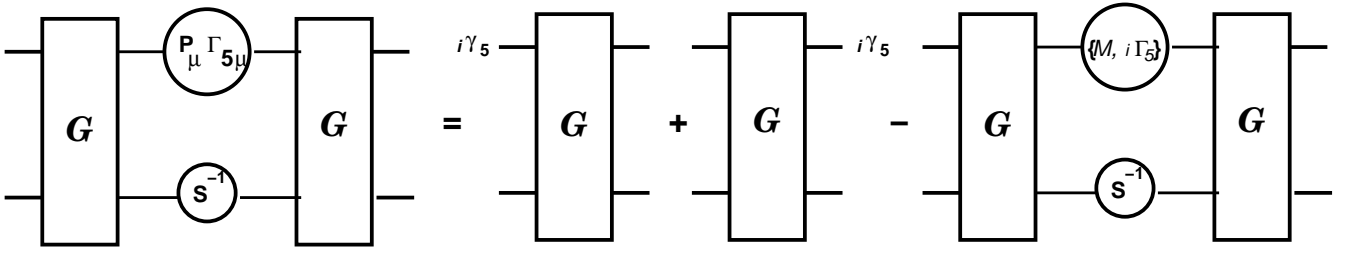}
\parbox{\textwidth}{\caption{\label{figAVWTI} This axial-vector Ward-Takahashi identity is an analogue of Eq.\,(\protect\ref{avwti0}).  It is valid if, and only if: the dressed-quark propagator, $S$, is obtained from Eq.\,(\ref{rainbowdse}); the axial-vector vertex, $\Gamma_{5\mu}$, is obtained from Eq.\,(\protect\ref{avbse}) with the kernel constructed from $S$ and Eq.\,(\protect\ref{ladderK}); the pseudoscalar vertex is constructed analogously; and the unamputated renormalised quark-antiquark scattering matrix: $G= (SS) + (SS)K(SS) + (SS)K(SS)K(SS)+ [\ldots]$, is constructed from the elements just described.}}
\end{center}
\end{figure}
\end{widetext}

The dressed-quark propagators in Eqs.\,(\ref{Tmnr}) -- (\ref{PVregd}) are understood to be calculated using the rainbow-truncation gap equation, which is defined by Eq.\,(\ref{gendse}) with 
\begin{equation}
\Sigma(p)=\int^\Lambda_q\! {\cal G}((p-q)^2) D_{\mu\nu}^{\rm free}(p-q) \frac{\lambda^a}{2}\gamma_\mu S(q) \frac{\lambda^a}{2}\gamma_\nu , \label{rainbowdse} 
\end{equation} 
wherein $D_{\mu\nu}^{\rm free}(\ell)$ is the free gauge boson propagator \cite{fn:landau} and ${\cal G}(\ell^2)$ will subsequently be specified.  The remaining element, the axial-vector vertex, is obtained from the ladder Bethe-Salpeter equation, whose kernel (see Eq.\,(\ref{bse1}), for example) is defined by the dressed-quark propagators just specified and  
\begin{eqnarray}
\nonumber \lefteqn{
K^{tu}_{rs}(q,k;P) =  }\\
&& \!\!\! - \,{\cal G}((k-q)^2) \, D_{\mu\nu}^{\rm free}(k-q)\,\left[\gamma_\mu \frac{\lambda^a}{2}\right]_{ts} \, \left[\gamma_\nu \frac{\lambda^a}{2}\right]_{ru} \!\!\!. \label{ladderK}
\end{eqnarray}

In what follows it is important that the rainbow-ladder truncation is the first term in the systematic and symmetry preserving truncation scheme described in Refs.\,\cite{bender,detmold,mandarvertex} and, furthermore, that with the choice 
\begin{equation}
\label{calGuv}
{\cal G}(\ell^2) = 4\pi \alpha_S(\ell^2)\,,\; \ell^2\gg \Lambda_{\rm QCD}^2\,,
\end{equation}
the rainbow-ladder truncation is guaranteed to express the one-loop renormalisation group properties of QCD. 

The axial-vector Ward-Takahashi identity depicted in Fig.\,\ref{figAVWTI} is an analogue of
\begin{eqnarray}
\nonumber \lefteqn{P_\mu {\cal S}(k_+) \,\Gamma_{5\mu}^j(k;P)\, {\cal S}(k_-)  = i \gamma_5\frac{\tau^j}{2} \, {\cal S}(k_-)   }\\
\nonumber 
&+&   {\cal S}(k_+)\, i \gamma_5\frac{\tau^j}{2}  - {\cal S}(k_+) \{{\cal M}(\zeta)\, , \,i\Gamma_5^j(k;P)\} {\cal S}(k_-)\,.\\
\label{avwti0}
\end{eqnarray}
It can be derived following the method in Refs.\,\cite{bicudo,marisbicudo} if, and only if, every dressed-quark propagator that appears is obtained from the rainbow DSE and the accompanying dressed vertices are determined from the ladder Bethe-Salpeter equation, both of which have just been defined.  

Using the identity in Fig.\,\ref{figAVWTI} it can be shown \cite{lc03} that 
\begin{equation}
\label{anomaly}
P_\rho T^3_{5\mu\nu\rho}(k_1,k_2) + 2 i m(\zeta) \, T^3_{5\mu\nu}(k_1,k_2) = \frac{\alpha}{2 \pi} \varepsilon_{\mu\nu\rho\sigma} k_{1\rho} k_{2\sigma}\,,
\end{equation}
where $\alpha= e^2/(4\pi)$.  This is an explicit demonstration that the triangle-diagram representation of the axial-vector--two-photon coupling calculated in the rainbow-ladder truncation is a necessary and sufficient pairing to preserve the Abelian anomaly.

In general the coupling of an axial-vector current to two photons is described by a six-point Schwinger function, to which Eq.\,(\ref{Tmnr}) is an approximation.  The same is true of the pseudoscalar--two-photon coupling and its connection with Eq.\,(\ref{Pmnr}).  Equation~(\ref{anomaly}) is valid for any and all values of $P^2=(k_1+k_2)^2$.  It is an exact statement of a divergence relation between these two six-point Schwinger functions, which is preserved by the truncation we will subsequently employ in illustrative quantitative studies.  Before providing those illustrations, however, we derive corollaries of Eq.\,(\ref{anomaly}) that have important implications for the properties of pseudoscalar bound states.

If one inserts Eqs.\,(\ref{genavv}) and (\ref{genpv}) into Eq.\,(\ref{anomaly}) and uses Eq.\,(\ref{gmorgen}), one finds that in the neighbourhood of each electric-charge-neutral pseudoscalar-meson bound-state pole  
\begin{eqnarray}
\nonumber \lefteqn{P_\rho T_{5\mu\nu\rho}^{3\,{\rm reg}}(k_1,k_2) + 2 i m(\zeta) \, T_{5\mu\nu}^{3\,{\rm reg}}(k_1,k_2)}\\
& &   + f_{\pi_n} \,T^{\pi_n^0}_{\mu\nu}(k_1,k_2) = \frac{\alpha}{2 \pi} i\varepsilon_{\mu\nu\rho\sigma} k_{1\rho} k_{2\sigma}\,.
\label{reganomaly}
\end{eqnarray}
In this equation, $T^{3\,{\rm reg}}(k_1,k_2)$ are nonresonant or \emph{continuum} contributions to the relevant Schwinger functions, whose form is concretely illustrated herein upon substitution of $\Gamma_{5 \mu}^{j\,{\rm reg}}(k;P)$ and $\Gamma_{5}^{j\,{\rm reg}}(k;P)$ into Eqs.\,(\ref{Tmnr}) and (\ref{Pmnr}), respectively.  Moreover, $T^{\pi_n^0}$ is the six-point Schwinger function describing the bound state contribution, which in rainbow-ladder truncation is realised as   
\begin{eqnarray}
\nonumber T^{\pi_n^0}_{\mu\nu}(k_1,k_2)  &=& {\rm tr}\int_\ell^{M\to\infty} \!\! {\cal S}(\ell_{0+}) \, \Gamma_{\pi_n^0}(\ell_{-\frac{1}{2}\frac{1}{2}};P) \, {\cal S}(\ell_{-0}) \\%
&\times&  \, i{\cal Q}\Gamma_\mu(\ell_{-0},\ell) \, {\cal S}(\ell) \, i {\cal Q}\Gamma_\nu(\ell,\ell_{0+}).
\label{Tpingg}
\end{eqnarray}
This Schwinger function describes the direct coupling of a pseudoscalar meson to two photons.  The support properties of the bound state Bethe-Salpeter amplitude guarantee that the renormalised Schwinger function is finite so that the regularising parameter can be removed; i.e., $M\to \infty$, in general and in our truncation, Eq.\,(\ref{Tpingg}).  

We note that owing to the $O(4)$ (Euclidean Lorentz) transformation properties of each term on the l.h.s.\ in Eq.\,(\ref{anomaly}), one may write
\begin{eqnarray}
P_\rho T_{5\mu\nu\rho}^{3\,{\rm reg}}(k_1,k_2) & = & \frac{\alpha}{\pi} i\varepsilon_{\mu\nu\rho\sigma} k_{1\rho} k_{2\sigma}\,A^{3\,{\rm reg}}(k_1,k_2) \,,\; \\
T_{5\mu\nu}^{3\,{\rm reg}}(k_1,k_2) & = & \frac{\alpha}{\pi} i\varepsilon_{\mu\nu\rho\sigma} k_{1\rho} k_{2\sigma}\,P^{3\,{\rm reg}}(k_1,k_2)\,,\; \\
T^{\pi_n^0}_{\mu\nu}(k_1,k_2) & = & \frac{\alpha}{\pi} i\varepsilon_{\mu\nu\rho\sigma} k_{1\rho} k_{2\sigma}\,G^{\pi_n^0}(k_1,k_2)\,, \; \label{TGdef}
\end{eqnarray}
so that Eq.\,(\ref{anomaly}) can be compactly expressed as
\begin{equation}
\label{reganomaly0}
A^{3\,{\rm reg}}(k_1,k_2) + 2 i m(\zeta) P^{3\,{\rm reg}}(k_1,k_2) + f_{\pi_n} G^{\pi_n^0}(k_1,k_2) = \frac{1}{2}.
\end{equation}

It has been proven \cite{andreasrapid} that in the chiral limit
\begin{equation}
\label{fpizero}
f_{\pi_n}^0 \equiv 0\; \forall n\geq 1.
\end{equation}  
Hence it follows from Eq.\,(\ref{reganomaly}) that in this limit all pseudoscalar mesons, \emph{except} the Goldstone mode, decouple from the divergence of the axial-vector--two-photon vertex. (This is true unless $G^{\pi_n^0}(k_1,k_2)$ diverges in the chiral limit, which is not the case, as we will see.)  

In the chiral limit the pole associated with the ground state pion appears at $P^2=0$ and thus  
\begin{eqnarray}
\nonumber
\lefteqn{\left. P_\rho T_{5\mu\nu\rho}^{3}(k_1,k_2)\right|_{P^2\neq 0}}\\
&&  = \left.P_\rho T_{5\mu\nu\rho}^{3\,{\rm reg}}(k_1,k_2)\right|_{P^2\neq 0} = \frac{\alpha}{2 \pi} i\varepsilon_{\mu\nu\rho\sigma} k_{1\rho} k_{2\sigma}\,;
\end{eqnarray}
namely, outside the neighbourhood of the ground state pole the regular (or continuum) part of the divergence of the axial-vector vertex saturates the anomaly in the divergence of the axial-vector--two-photon coupling.  

On the other hand, in the neighbourhood of $P^2=0$ 
\begin{eqnarray} 
\left. A^{3\,{\rm reg}}(k_1,k_2) \right|_{ P^2\simeq 0} + f_{\pi_0} \,G^{\pi_0}(k_1,k_2)
& =& \frac{1}{2}\,; \label{anomalypion}
\end{eqnarray}
i.e., on this domain the contribution to the axial-vector--two-photon coupling from the regular part of the divergence of the axial-vector vertex combines with the direct $\pi_0^0 \gamma \gamma$ vertex to fulfill the anomaly.  This fact was illustrated in Ref.\,\cite{mrpion} by direct calculation: Eqs.\,(\ref{bwti}) -- (\ref{hwti}) are an essential part of that demonstration.

If one defines 
\begin{equation}
\label{TpiG}
{\cal T}_{\pi_n^0}(P^2,Q^2) = \left. G^{\pi_n^0}(k_1,k_2) \right|_{k_1^2=Q^2=k_2^2},
\end{equation}
in which case $ P^2= 2(k_1\cdot k_2+Q^2)$, then the physical width of the neutral ground state pion is determined by
\begin{equation}
g_{\pi_0^0 \gamma\gamma}:= {\cal T}_{\pi_0^0}(-m_{\pi_0^0}^2,0) ;
\end{equation}
viz., the second term on the l.h.s.\ of Eq.\,(\ref{anomalypion}) evaluated at the on-shell points.  This result is not useful unless one has a means of estimating the contribution from the first term; viz., $A^{3\,{\rm reg}}(k_1,k_2)$.  However, that is readily done.  A consideration \cite{mrt98} of the structure of the regular piece in Eq.\,(\ref{genavv}) indicates that the impact of this continuum term on the $\pi_0^0 \gamma\gamma$ coupling is modulated by the magnitude of the pion's mass, which is small for realistic $u$ and $d$ current-quark masses and vanishes in the chiral limit.  One therefore expects this term to contribute very little and anticipates from Eq.\,(\ref{anomalypion}) that 
\begin{equation}
\label{anomalycouple}
g_{\pi_0^0 \gamma\gamma} = \frac{1}{2} \frac{1}{f_{\pi_0}}
\end{equation}
is a good approximation.  This is verified in explicit calculations; e.g., in Ref.\,\cite{maristandypi0}, which evaluates the triangle diagrams described herein, the first term on the l.h.s.\ modifies the result in Eq.\,(\ref{anomalycouple}) by less than 2\%.  

There is no reason to expect an analogous result for pseudoscalar mesons other than the $\pi(140)$; i.e., the states which we denote by $n\geq 1$.  Indeed, as all known such pseudoscalar mesons have experimentally determined masses that are greater than $1\,$GeV, the reasoning used above suggests that the presence of the continuum terms, $A^{3\,{\rm reg}}(k_1,k_2)$ and $P^{3\,{\rm reg}}(k_1,k_2)$, must materially impact upon the value of $g_{\pi_n^0 \gamma\gamma}$.  This will subsequently be illustrated using the rainbow-ladder truncation.

\subsection{Asymptotic behaviour of transition form factor}
\label{exactUV}
We have stated that the rainbow-ladder truncation preserves the one-loop renormalisation group properties of QCD.  It follows that Eq.\,(\ref{Tpingg}) should reproduce the leading large-$Q^2$ behaviour of the $\gamma^\ast(Q) \pi_n(P) \gamma^\ast(Q)$ transition form factor inferred from perturbative QCD.  The QCD analysis has been performed for the ground state pion ($n=0$) with the result \cite{uvQQ}
\begin{equation}
\label{TpiuvQCD}
{\cal T}_{\pi_0^0}(P^2=-m_{\pi_0}^2,Q^2) \stackrel{Q^2\gg \Lambda_{\rm QCD}^2}{=} \frac{4\pi^2}{3} \frac{f_{\pi_0}}{Q^2}\,,
\end{equation}
and Ref.\,\cite{kekez} verified that this is indeed the result contained in Eq.\,(\ref{Tpingg}).  However, it is useful for our purposes to recapitulate on that derivation.

Consider Eq.\,(\ref{Tpingg}): the integral is finite and hence a shift in the integration variable is permitted, 
\begin{eqnarray}
\nonumber \lefteqn{T^{\pi_n^0}_{\mu\nu}(k_1,k_2)  = {\rm tr}\int_\ell^{M\to \infty} \!\! \chi_{\pi_n^0}(\ell;P) }\\%
\nonumber &\times&  i{\cal Q} \Gamma_\mu(\ell_{-P},\ell_{K}) \, {\cal S}(\ell_{K}) \, i {\cal Q}\Gamma_\nu(\ell_{K},\ell_{P}),\\
\label{TpinggN}
\end{eqnarray}
where $\ell_P:= \ell_{\frac{1}{2}\frac{1}{2}}= \ell + P/2$ and $\ell_K:= \ell_{\frac{1}{2}-\frac{1}{2}}=: \ell + K$.  We assume that $k_1^2=Q^2=k_2^2$ with $Q^2\gg \Lambda_{\rm QCD}^2$ and, because we do not restrict ourselves to ground state pseudoscalar mesons, assume besides that for the given $n$ under consideration $Q^2 \gg m_{\pi_n}^2$.  On this domain $K\cdot P\equiv 0$, $K^2=Q^2$, and it is valid at leading $(1/Q^2)$-order in Eq.\,(\ref{TpinggN}) to write \cite{cdrcroat,pctcroat}
\begin{eqnarray}
\label{expand}
\nonumber &&  i{\cal Q} \Gamma_\mu(\ell_{-P},\ell_{K}) \, {\cal S}(\ell_{K}) \, i {\cal Q}\Gamma_\nu(\ell_{K},\ell_{P}) \\
& = & Z_2 \, i{\cal Q} \gamma_\mu \, \frac{-i\gamma\cdot \ell_{K}}{\ell_{K}^2} \, i{\cal Q} \gamma_\nu
\end{eqnarray}
so that
\begin{eqnarray}
\nonumber \lefteqn{ T^{\pi_n^0}_{\mu\nu}(k_1,k_2)    }\\
\nonumber &=&  \frac{4 \pi \alpha}{3}\, i\varepsilon_{\mu\nu\rho\sigma}\, {\rm tr} \, Z_2 \int_\ell^{M} \!\! \sfrac{1}{2} \tau^3\, \gamma_5 \gamma_\sigma \, \chi_{\pi_n^0}(\ell;P)  \,  \frac{(\ell_{K})_\rho}{\ell_{K}^2}.\\
\label{TpinggNa}
\end{eqnarray}

Since we are concerned with $J^{PC} = 0^{-+}$ states, it follows that 
\begin{eqnarray}
\nonumber \lefteqn{ T^{\pi_n^0}_{\mu\nu}(k_1,k_2)  =  \frac{4 \pi \alpha}{3}\, i\varepsilon_{\mu\nu\rho\sigma}}\\
& & \times  \left[K_\rho {\cal I}_\sigma(K,P) - K_\alpha {\cal J}_{\rho\sigma\alpha}(K,P)\right],
\label{Tanswer}
\end{eqnarray}
where Eq.\,(\ref{TpinggNa}) yields
\begin{eqnarray}
\nonumber
\lefteqn{{\cal I}_\sigma(K,P) }\\
\nonumber
&= & {\rm tr} \, Z_2 \int_\ell^{M} \!\! \sfrac{1}{2} \tau^3\, \gamma_5 \gamma_\sigma \, \chi_{\pi_n^0}(\ell;P) \,(\ell^2 + K^2) \, \Delta(\ell,K)  \\ \label{Ires}\\
\nonumber
\lefteqn{K_\alpha {\cal J}_{\rho\sigma\alpha}(K,P)}\\
\nonumber
&=&  {\rm tr} \, Z_2 \int_\ell^{M} \!\! \sfrac{1}{2} \tau^3\, \gamma_5 \gamma_\sigma \, \chi_{\pi_n^0}(\ell;P)   \, 2 \, \ell_\rho\, \ell\cdot K \, \Delta(\ell,K)\\
\label{Jres}
\end{eqnarray}
with $\Delta(l,K) = 1/[(\ell^2+K^2)^2 - 4 (\ell\cdot K)^2]$.

As we show in the Appendix, on the large-$Q^2$ domain, that part of ${\cal I}_\sigma(K,P)$ which contributes to $T^{\pi_n^0}_{\mu\nu}(k_1,k_2)$ is
\begin{equation}
{\cal I}_\sigma(K,P) = P_\sigma 
\left\{ \frac{f_{\pi_n}}{Q^2} + F^{(2)}_{\cal I}(P^2) \frac{\ln^{\gamma} Q^2/\omega_{\pi_n}^2}{Q^4} \right\}, \label{Iuv}
\end{equation}
$P^2= -m_{\pi_n}^2$, where $\gamma$ is an anomalous dimension and $\omega_{\pi_n}$ is a mass-scale associated with the momentum space width of the meson's Bethe-Salpeter wave function.  Similar reasoning exposes the leading contribution to Eq.\,(\ref{Tanswer}) from Eq.\,(\ref{Jres}):
\begin{equation}
K_\alpha {\cal J}_{\rho\sigma\alpha}(K,P) = K_\rho P_\sigma \, F^{(2)}_{\cal J}(P^2)\frac{\ln^{\gamma} Q^2/\omega_{\pi_n}^2}{Q^4} \,,
\end{equation}
$P^2= -m_{\pi_n}^2$.  Combining these results one arrives at
\begin{eqnarray}
\nonumber \lefteqn{
 T^{\pi_n^0}_{\mu\nu}(k_1,k_2)  \stackrel{Q^2\to \infty}{=} \frac{4 \pi \alpha }{3} i\varepsilon_{\mu\nu\rho\sigma}\, k_{1\rho} k_{2\sigma}  }\\
 &\times &  \left[\frac{f_{\pi_n}}{Q^2} + F^{(2)}_{n }(P^2)\frac{\ln^{\gamma} Q^2/\omega_{\pi_n}^2}{Q^4} \right].\label{enduv}
\end{eqnarray} 

We emphasise that the coefficient of the leading $1/Q^2$-term in Eq.\,(\ref{enduv}) is exact and model-independent.  

That is not true of the subleading $1/Q^4$ term.  Furthermore, with a given \textit{Ansatz} for ${\cal G}(k^2)$ in Eqs.\,(\ref{rainbowdse}) and (\ref{ladderK}), Eq.\,(\ref{expand}) is not sufficient to accurately determine the value of the coefficient of the $1/Q^4$ term or the anomalous dimension because, for example, momentum-dependent dressing of the quark-photon vertex can contribute at this order.  Nevertheless, our analysis highlights the existence of a nonzero subleading $1/Q^4$ contribution whose strength is sensitive to features of the dynamics.  These observations were made previously for the ground state ($n=0$) pion \cite{yeh}.

We can now return to one of the stated reasons for this analysis: Eq.\,(\ref{enduv}) inserted in Eq.\,(\ref{TGdef}) and combined with Eq.\,(\ref{TpiG}) reproduces the leading order result obtained in perturbative QCD, Eq.\,(\ref{TpiuvQCD}).  In fact, it provides more.  The perturbative result was only derived for the ground state pseudoscalar meson.  Our analysis shows that for each meson on the pseudoscalar trajectory, identified herein by a value of $n$, QCD predicts
\begin{eqnarray}
\nonumber
\lefteqn{{\cal T}_{\pi_n^0}(-m_{\pi_n}^2,Q^2) \stackrel{Q^2\gg \Lambda_{\rm QCD}^2}{=} \frac{4\pi^2}{3}}\\
& \times & \left[ \frac{f_{\pi_n}}{Q^2} + F_n^{(2)}(-m_{\pi_n}^2) 
\frac{\ln^{\gamma} Q^2/\omega_{\pi_n}^2}{Q^4}
\right] .
\label{UVnot0}
\end{eqnarray}

It is now apparent from Eq.\,(\ref{fpizero}) that $\forall n\geq 1$ 
\begin{eqnarray}
\nonumber \lefteqn{\lim_{\hat m\to 0} {\cal T}_{\pi_n^0}(-m_{\pi_n}^2,Q^2) }\\
&& \stackrel{Q^2\gg \Lambda_{\rm QCD}^2}{=} \frac{4\pi^2}{3}\left. F^{(2)}_{n }(-m_{\pi_n}^2)\frac{\ln^{\gamma} Q^2/\omega_{\pi_n}^2}{Q^4}\right|_{\hat m=0} \,;
\label{UVchiralnot0}
\end{eqnarray}
namely, in the chiral limit the leading-order power-law in the transition form factor for excited state pseudoscalar mesons is O$(1/Q^4)$.  This result is model-independent.  

Furthermore, while we cannot determine the QCD value of the coefficient $F_n^{(2)}(-m_{\pi_n}^2)$ in the present truncation, in general that coefficient is \emph{not} proportional to $f_{\pi_n}$, or some power thereof, for any value of $n$.  We will see this clearly in the $n\geq 1$ transition form factor for which,  if that were the case, the $1/Q^4$-term would be absent in the chiral limit.  For all pseudoscalar states there are mass-scales other than $f_\pi$ that are nonzero even in the chiral limit when chiral symmetry is dynamically broken.  

\section{Couplings of Pseudoscalar Mesons: Model Results}
\label{model}
\subsection{Rainbow-ladder truncation}
\label{sec:rl}
In order to illustrate the results presented above and calculate other observables it is necessary to specify ${\cal G}(k^2)$ in Eqs.\,(\ref{rainbowdse}) and (\ref{ladderK}).  We choose 
\begin{equation}
\label{calG}
\frac{{\cal G}(s)}{s} = \frac{4\pi^2}{\omega^6} \, D\, s\, {\rm e}^{-s/\omega^2}+ \frac{8\pi^2 \gamma_m}{\ln\left[ \tau + \left(1+s/\Lambda_{\rm QCD}^2\right)^2\right]} \, {\cal F}(s)\,,
\end{equation}
with ${\cal F}(s)= [1-\exp(-s/[4 m_t^2])]/s$, $m_t=0.5\,$GeV, $\ln(\tau+1)=2$, $\gamma_m=12/25$ and $\Lambda_{\rm QCD} = \Lambda^{(4)}_{\overline{MS}} = 0.234\,$GeV.  

This form expresses the interaction as a sum of two terms.  The second guarantees Eq.\,(\ref{calGuv}) and therefore ensures that perturbative behaviour is correctly realised at short range; namely, as written, for $(k-q)^2 \sim k^2 \sim q^2 \gtrsim 1 - 2\,$GeV$^2$, $K$ is precisely as prescribed by QCD.  On the other hand, the first term in ${\cal G}(k^2)$ is a model for the long-range behaviour of the interaction.  It is a finite width representation of the form introduced in Ref.\,\cite{mn83}, which has been rendered as an integrable regularisation of $1/k^4$ \cite{mm97}.  This interpretation, when combined with the result that in a heavy-quark--heavy-antiquark BSE the renormalisation-group-improved ladder truncation is exact \cite{mandarvertex}, is consistent with ${\cal G}(k^2)$ leading to a Richardson-like potential \cite{richardson} between static sources.  

The active parameters in Eq.\,(\ref{calG}) are $D$ and $\omega$, which together determine the integrated infrared strength of the rainbow-ladder kernel, but they are not independent.  In fitting a selection of ground state observables \cite{mt99}, a change in one is compensated by altering the other; e.g., on the domain $\omega\in[0.3,0.5]\,$GeV, the fitted observables are approximately constant along the trajectory 
\begin{equation}
\omega D  = (0.72 \, {\rm GeV})^3 =: m_g^3\,.
\end{equation}  
(NB. The value of $m_g$ is typical of the mass-scale associated with nonperturbative gluon dynamics.)  Herein, unless otherwise stated, we use 
\begin{equation}
\label{omegavalue}
\omega= 0.35\,{\rm GeV.}
\end{equation}

Equation~(\ref{calG}) defines a renormalisation-group-im\-proved rainbow-ladder truncation.  This form, introduced in Refs.\,\cite{mr97,mt99}, has been employed extensively in the calculation of properties of ground state pseudoscalar and vector mesons \cite{fn:jain}.  These applications are reviewed in Ref.\,\cite{pieterrev}, from which it is apparent that the model describes a basket of thirty-one hadron observables with a rms error between calculation and experiment of $15$\%.  

The calculation of observables is now straightforward.  The kernel of the gap equation, Eq.\,(\ref{rainbowdse}), is completely specified.  Thus a solution follows immediately upon fixing the current-quark mass: this sets the \emph{boundary condition}, Eq.\,(\ref{renormS}).  We focus on the $u$-$d$ sector and assume isospin symmetry: 
\begin{equation}
\hat m_u=\hat m_d= \hat m\,.
\end{equation}  
With a result for the dressed-quark propagator in hand, the kernel of Bethe-Salpeter equations is also complete.  The solutions of these equations yield: the bound state Bethe-Salpeter amplitudes; the axial-vector and pseudoscalar vertices; and the dressed-quark-photon vertex, all of which appear above.  At this point one has every element necessary for the calculation of an amplitude such as Eq.\,(\ref{TpinggN}) and therewith experimental observables.  The numerical procedures are described in Refs.\,\cite{mr97,mt99,mt00,krassnigg1}.

\begin{figure}[t]
\begin{center}
\includegraphics[width=0.48\textwidth]{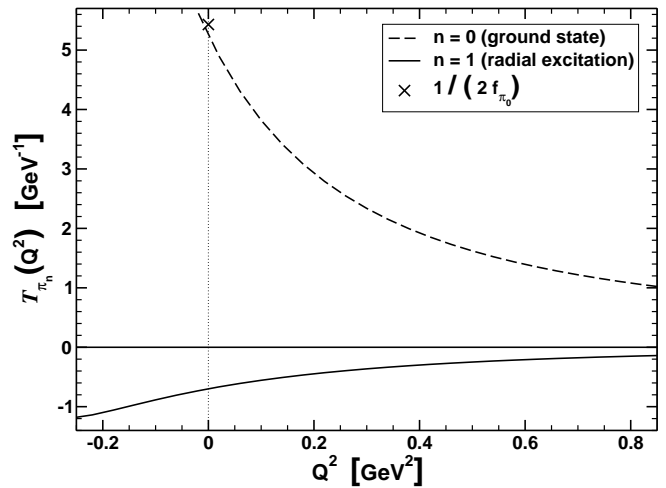}
\caption{\label{fig:Tpiggnmm} Small-$Q^2$ behaviour of the $\gamma^\ast(Q) \pi_n(P) \gamma^\ast(Q)$ transition form factor, defined in Eq.\,(\protect\ref{TpiG}), calculated with the current-quark mass in Eq.\,(\protect\ref{qmass}).  The ground state's two-photon coupling suggested by Eq.\,(\protect\ref{anomalycouple}) is marked by ``$\times$''. }
\end{center}
\end{figure}

\subsection{Two photon couplings of pseudoscalar mesons}
Figure~\ref{fig:Tpiggnmm} depicts the small-$Q^2$ behaviour of the $\gamma^\ast(Q) \pi_n(P) \gamma^\ast(Q)$ transition form factor defined in Eq.\,(\protect\ref{TpiG}), calculated for the two lowest-mass $0^{-+}$ states with 
\begin{equation}
\label{qmass}
m(\zeta_0) := \frac{\hat m}{(\ln\zeta_0/\Lambda_{\rm QCD})^{\gamma_m}} = 5.5\,{\rm MeV}\,,\; \zeta_0=  1\,{\rm GeV}\,.
\end{equation}
(Recall that in this model the $n=1$ state is a radial excitation.)  It is notable that while ${\cal T}_{\pi_0^0}(-m_{\pi_0}^2,Q^2)>0$, 
\begin{equation}
{\cal T}_{\pi_1^0}(-m_{\pi_1}^2,Q^2)<0 \,, \; Q^2\geq -m_{\pi_1}^2/4 ;
\end{equation}
viz., it is negative on the entire kinematically accessible domain.  Moreover, for nonzero current-quark mass we expect the sign of this form factor to duplicate the pattern set by the leptonic decay constant, which is $(-1)^n$ \cite{andreasrapid}.  NB.\ On the depicted domain and with the resolution in this figure there is no perceptible difference between these curves and those obtained in the chiral limit.  That is not true for larger $Q^2$, as will become apparent.

The coupling constants for decay into two real photons are presented in Table~\ref{table:couplings}, as are the associated decay widths, calculated using
\begin{equation}
\label{ggwidth}
\Gamma_{\pi^0_n \gamma\gamma} = \alpha_{\rm em}^2\, \frac{m_{\pi_n}^3}{16\pi^3} \,  g^2_{\pi_n \gamma\gamma}.
\end{equation}
It is evident from Table~\ref{table:couplings} that Eq.\,(\protect\ref{anomalycouple}) is truly a good approximation for the $\pi(140)$.  

\begin{table}[t] 
\caption{\label{table:couplings} Results for a range of properties of the two lowest mass $0^{-+}$ mesons.  Note that for $n=0$, Eq.\,(\protect\ref{anomalycouple}) yields: chiral limit, $5.68\,$GeV$^{-1}$; massive, Eq.\,(\protect\ref{qmass}), $5.41\,$GeV$^{-1}$.  Decay widths: calculated from Eqs.\,(\protect\ref{ggwidth}); value known experimentally \cite{pdg}: $\Gamma_{\pi_0 \gamma\gamma}=7.84\pm 0.56\,$eV. Also \protect\cite{pdg}: $m_{\pi_0} = 0.14\,$GeV; $m_{\pi_1} = 1.3\pm 0.1\,$GeV.  [NB.\ Our best estimate is $\Gamma_{\pi_1^0 \gamma\gamma} \approx 240$eV, for reasons presented in connection with Eq.\,(\protect\ref{Gpiggbest}).]}
\begin{ruledtabular} 
\begin{tabular*} 
{\hsize} {l@{\extracolsep{0ptplus1fil}} 
l@{\extracolsep{0ptplus1fil}}|l@{\extracolsep{0ptplus1fil}}
l@{\extracolsep{0ptplus1fil}}l@{\extracolsep{0ptplus1fil}}l@{\extracolsep{0ptplus1fil}}}
 & & $m_n\,$ & $f_n\,$ & $g_{{\pi_n}\gamma\gamma}$ & $\Gamma_{\pi_n^0 \gamma\gamma}$ \\
 & & (GeV) & (GeV) & (GeV)$^{-1}$ & (eV) \\\hline
$\pi_0$ & $\hat m =0$ & $0.0$ & $\;\;\;0.088$ & $\;\;\;5.31$ & \\
& $\hat m$, Eq.\,(\protect\ref{qmass})~ & $0.14$ & $\;\;\;0.092$ & $\;\;\;5.25$ & $\;\;7.9$ \\
$\pi_1$ & $\hat m =0$ & $1.04$ & $\;\;\;0.0$ & $-0.71$ \\
& $\hat m$, Eq.\,(\protect\ref{qmass})~ & $1.06$ & $-0.0016$ & $-0.70$ & $63.0$\\
\end{tabular*} 
\end{ruledtabular} 
\end{table} 

The result for $g_{\pi_1 \gamma\gamma}$ is, however, striking.  This coupling is negative because the $\pi_1$'s Bethe-Salpeter amplitude has a significant domain of negative support \cite{andreasrapid}; and while its magnitude is material, $\sim 0.13\,g_{\pi_0 \gamma\gamma}$, it is finite even in the chiral limit.  The last fact demonstrates that the $\pi_1 \gamma\gamma$ coupling is not inversely proportional to $f_{\pi_1}$ cf.\ Eq.\,(\ref{anomalycouple}).  This confirms that the excited state decouples from the axial-vector--two-photon vertex in the chiral limit, as described in connection with Eq.\,(\ref{reganomaly0}).  Consequently, the evolution with $P^2$ of the regular (or continuum) part of the divergence of the axial-vector--two-photon vertex is smooth; i.e.,  
\begin{equation}
 \left. A^{3\,{\rm reg}}(k_1,k_2) \right|_{ P^2\simeq   -m_{\pi_1}^2} \approx \left. A^{3\,{\rm reg}}(k_1,k_2) \right|_{ P^2= -m_{\pi_1}^2} \,,
\end{equation}
and in addition
\begin{equation}
\left[A^{3\,{\rm reg}}(k_1,k_2) + 2 i m(\zeta) P^{3\,{\rm reg}}(k_1,k_2) \right]_{ P^2= -m_{\pi_1}^2} \approx \frac{1}{2}\,,
\end{equation}
with exact equality for $\hat m =0$.

\begin{figure}[t]
\begin{center}
\includegraphics[width=0.48\textwidth]{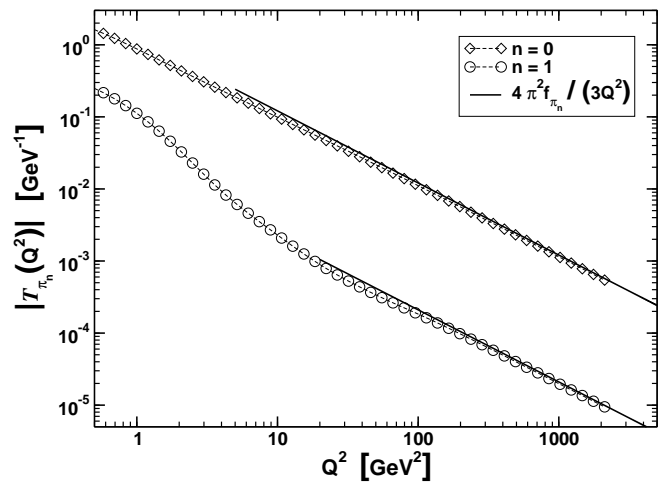}
\caption{\label{fig:UV01m} Calculated large-$Q^2$ behaviour of the $\gamma^\ast(Q) \pi_n(P) \gamma^\ast(Q)$ transition form factor, Eq.\,(\protect\ref{TpiG}): \textit{diamonds} -- ground state, $n=0$; and \textit{circles} -- first excited state, $n=1$.   The \textit{solid-lines} are Eq.\,(\protect\ref{TpiuvQCD}) with either $f_{\pi_0}$ or $f_{\pi_1}$ from Table~\protect\ref{table:couplings}, as appropriate.}
\end{center}
\end{figure}

In Fig.\,\ref{fig:UV01m} we depict the large-$Q^2$ behaviour of the $\gamma^\ast(Q) \pi_n(P) \gamma^\ast(Q)$ transition form factor obtained with the nonzero current-quark mass in Eq.\,(\ref{qmass}), for the two lowest mass pseudoscalars.  The ultraviolet behaviour anticipated for the ground state from perturbative QCD, Eq.\,(\ref{TpiuvQCD}), is evident.  This is a numerical verification of the argument associated with Eqs.\,(\ref{TpinggN}) -- (\ref{UVchiralnot0}); viz., that the truncation we employ preserves leading-order QCD results.  The analogous result for the first excited state, indicated by Eq.\,(\ref{UVnot0}), is also conspicuous. 

\begin{figure}[t]
\begin{center}
\includegraphics[width=0.48\textwidth]{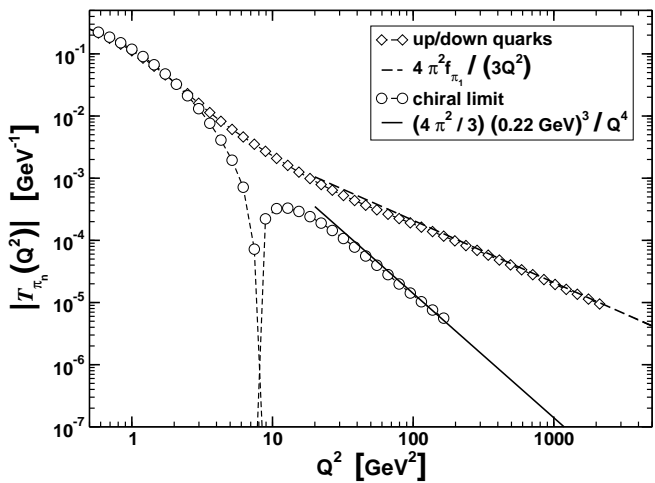}
\caption{\label{fig:UV1chiral} Large-$Q^2$ behaviour of the $\gamma^\ast(Q) \pi_1(P) \gamma^\ast(Q)$ transition form factor, Eq.\,(\protect\ref{TpiG}): \textit{Diamonds} -- the result obtained with $\hat m$ in Eq.\,(\protect\ref{qmass}); \textit{Circles} --  our chiral limit calculation ($\hat m = 0$); \textit{Solid line} -- the curve $\frac{4\pi^2}{3} (0.22\,{\rm GeV})^3/Q^4$.}
\end{center}
\end{figure}

For the ground state the behaviour of the transition form factor in the chiral limit is not markedly different from that found with $\hat m$ in Eq.\,(\ref{qmass}) and illustrated in Fig.\,\ref{fig:UV01m}.  As evident in Fig.\,\ref{fig:UV1chiral}, that is not the case for $\gamma^\ast(Q) \pi_1(P) \gamma^\ast(Q)$ in the chiral limit.  While the form factor is initially negative, as may be anticipated from Fig.\,\ref{fig:Tpiggnmm}, it is positive for $Q^2 \gtrsim 8\,$GeV$^2$ and the asymptotic behaviour indicated in Eq.\,(\ref{UVchiralnot0}) is exhibited for $Q^2\gtrsim 50\,$GeV$^2$.  With the model's parameter value specified in Eq.\,(\ref{omegavalue}), we find
\begin{equation}
\label{Fnvalue}
\left. F_1^{(2)}(-m_{\pi_1}^2) \, \ln^{\gamma} Q^2/\omega_{\pi_1}^2\right|_{\hat m=0} \approx (0.22\,{\rm GeV})^3.
\end{equation}
This mass-scale is commensurate with that set by the vacuum quark condensate.  The magnitude of $F_1^{(2)}$ depends on the model parameter.  So, too, does the precise location of the boundary between the domains on which the transition form factor has negative and positive support.  However, qualitative features, such as the existence of these domains, are robust.  

It is noteworthy that while $f_{\pi_1}\equiv 0$ algebraically in the chiral limit, in practice there is always a numerical error.  Hence, as is plain from Eq.\,(\ref{UVnot0}), there will inevitably be a value of $Q^2$ beyond which the erroneous nonzero value of $f_{\pi_1}$, produced by the numerical error, will come to dominate the chiral-limit transition form factor.  To obtain the value in Eq.\,(\ref{Fnvalue}) we estimated the magnitude of this pollution and subtracted it.   For this reason, within the accuracy of our numerical analysis, we cannot provide reliable information on the $\ln Q^2$-modification.  The figure hints, however, at the presence in our model of such a modification to the $1/Q^4$-behaviour.

\subsection{Charge radii}
At leading order in the truncation scheme we are using, and in the isospin symmetric limit, the elastic electromagnetic form factor of a pseudoscalar meson is described by 
\begin{eqnarray}
\nonumber 
\lefteqn{
 e\, (p_1+p_2) \, F_{\pi_n}(Q^2) := e \, \Lambda_{\mu}(p_1,p_2)}\\
\nonumber 
& = & {\rm tr} \int_\ell %
\chi_{\pi_n}(\ell_{0,\frac{1}{2}})\, 
i {\cal Q} \Gamma_{\mu}(\ell_{-\frac{1}{2}\frac{1}{2}},\ell_{\frac{1}{2}-\frac{1}{2}})\\
&& \times \, 
\chi_{\pi_n}(\ell_{\frac{1}{2}0};-p_2) \, 
{\cal S} (\ell_{\frac{1}{2}\frac{1}{2}})^{-1}\,, \label{piem}
\end{eqnarray}
with $Q=p_1-p_2$.  Each element that appears in the integrand is fully renormalised and the integral is finite.  The expression automatically satisfies \cite{mt00,cdrpion}
\begin{equation}
(p_1-p_2)_\mu \, \Lambda_{\mu}(p_1,p_2) = 0\,
\end{equation} 
and guarantees
\begin{equation}
F_{\pi_n}(Q^2=0) = 1\,.
\end{equation}
In Ref.\,\cite{mt00} the model described in Sec.\,\ref{sec:rl} was employed to calculate the electromagnetic form factor of the pion using Eq.\,(\ref{piem}).  The prediction was subsequently verified in a JLab experiment performed at intermediate $Q^2$ \cite{volmer}.

\begin{figure}[t]
\begin{center}
\includegraphics[width=0.48\textwidth]{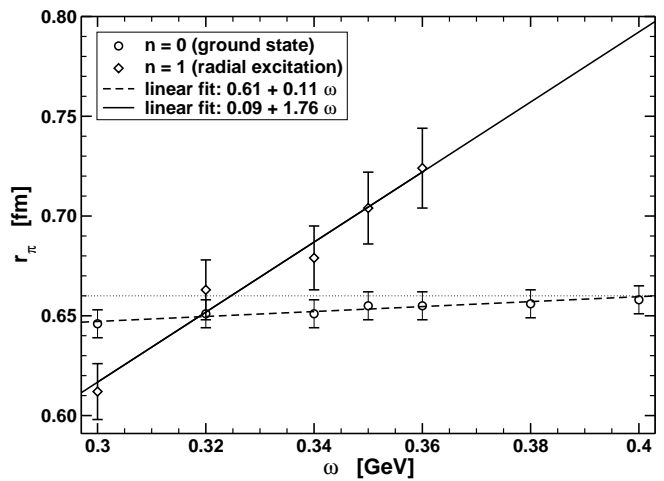}
\caption{\label{fig:emradii} Evolution of ground and first excited state pseudoscalar mesons' electromagnetic charge radii with the model's scale parameter $\omega$.  \textit{Dotted line}: $r_\pi=0.66\,$fm, which indicates the experimental value of the ground state's radius.  We must estimate the derivative in Eq.\,(\protect\ref{usualradius}) numerically.  That is the primary source of the numerical error depicted in the figure, which corresponds to a relative error  $\lesssim 1$\% for $n=0$ and $\lesssim 3$\% for $n=1$.}
\end{center}
\end{figure}

We have calculated the charge radii of the two lowest mass pseudoscalars using the standard definition:
\begin{equation}
\label{usualradius}
r_{\pi_n}^2 = - 6 \, F_{\pi_n}^\prime(Q^2=0)\,.
\end{equation}
Our results appear in Fig.\,\ref{fig:emradii}.  As promised in association with Eq.\,(\ref{omegavalue}), the ground state's properties are almost insensitive to the model's mass-scale, $\omega$: in formulating the model, a path appeared in the $(D,\omega)$ parameter space along which vacuum and ground state properties vary little.  The orthogonality of the excited states with respect to the ground state means there is no reason to expect such insensitivity in properties of the excited states.  And, indeed, one observes that the charge radius of the first excited state changes rapidly with increasing $\omega$, with the ratio $r_{\pi_1}/r_{\pi_0}$ varying from $0.9$ -- $1.2$.  

This outcome can readily be interpreted.  The length-scale $r_a := 1/\omega$ measures the range of strong attraction in our model: magnifying $r_a$ increases the range of strong attraction.  In Sec.\,\ref{sec:intro} we argued that the properties of radial excitations should be sensitive to the nature of the interaction between light quarks at long-range.  It is now apparent that this is true.  Moreover, decreasing $\omega$ has the effect of increasing the active range of the confining piece of the interaction in Eq.\,(\ref{calG}).  This effectively strengthens the confinement force.  That compresses the bound state, as one observes in Fig.\,\ref{fig:emradii}: $r_{\pi_1}$ decreases rapidly with decreasing $\omega$  (increasing $r_a$).  

A similar result for the evolution of the mass was observed in Ref.\,\cite{andreasrapid}; namely, the mass of the first excited state dropped rapidly with increasing $r_a$.  On the domain illustrated in Fig.\,\ref{fig:emradii}, the mass of the ground state obtained with nonzero current-quark mass varied by only 3\% while that of the first excited state changed by 14\%.  It is natural to expect that an increase in the strength of the confinement force should increase the magnitude of the binding energy and hence reduce the mass, and that is precisely what occurs.  (NB.\ Independent of the parameters, the ground state mass is identically zero in the chiral limit because the truncation is symmetry preserving.  Dynamical chiral symmetry breaking, which has many consequences, is another reason why properties of the ground state pseudoscalar meson do not respond rapidly to modest parameter changes.)   

It is natural to suppose $r_{\pi_1}>r_{\pi_0}$; namely, that a radial excitation is larger than the associated ground state.  However, our calculations illustrate that with the ground state pseudoscalar meson's properties constrained by Goldstone's theorem and its pointwise consequences, Eqs.\,(\ref{bwti}) -- (\ref{hwti}), it is possible for a confining interaction to compress the excited state with the consequence that $r_{\pi_1}<r_{\pi_0}$.  An analysis of the $\omega$-dependence of $m_{\pi_1}$ indicates that a value of $1.3\,$GeV may be obtained with $\omega \approx 0.48$ \cite{menu}.  However, quantitative difficulties connected with the behaviour of the dressed-quark propagator in the complex-$\ell^2$ plane \cite{jarecke,pichowskycomplex} currently prevent us from studying the excited state directly with $\omega > 0.4$ in Eq.\,(\ref{calG}).  Hence, we cannot make a firm prediction for $r_{\pi_1}$.  However, our results suggest $1.1 < r_{\pi_1}/r_{\pi_0}<1.6$, with a linear extrapolation giving
\begin{equation}
 r_{\pi_1}\simeq 1.4\,r_{\pi_0}\,.
\end{equation}

Naturally, we have also studied the evolution of $g_{\pi_n \gamma \gamma}$ with $\omega$.  On the domain illustrated in Fig.\,\ref{fig:emradii}, $g_{\pi_0 \gamma \gamma}$ varies by no more than 1\%, whereas $ g_{\pi_1 \gamma \gamma}(\omega=0.3)=-0.55$ and $ g_{\pi_1 \gamma \gamma}(\omega=0.4)=-0.80$, which is a variation over a range of $\sim 40$\%.  Following the reasoning above, and taking account of the variation in $m_{\pi_1}$, we conclude that it is likely $\Gamma_{\pi_1\gamma\gamma} > 150\,$eV $\gtrsim 20\,\Gamma_{\pi_0\gamma\gamma}$.  Our best estimate is $200 <\Gamma_{\pi_1\gamma\gamma} ({\rm eV}) <300$ and linear extrapolation gives
\begin{equation}
\label{Gpiggbest}
\Gamma_{\pi_1\gamma\gamma} \simeq 240\,{\rm eV}.
\end{equation}

\section{Epilogue}
\label{epilogue}
The strong interaction spectrum exhibits trajectories of mesons with the same spin\,$+$\,parity, $J^P$.  One may distinguish between the states on these trajectories by introducing an integer label $n$, with $n=0$ denoting the lowest-mass state, $n=1$ the next-lightest state, etc.  The Bethe-Salpeter equation (BSE) yields the mass and amplitude of every bound state in a given channel specified by $J^{P}$.  Hence it provides a practical tool for the Poincar\'e covariant study of mesons on these trajectories.  

In applying the Bethe-Salpeter equation to a study of pseudoscalar mesons we made use of the fact that at least one nonperturbative and symmetry preserving Dyson-Schwinger equation (DSE) truncation scheme exists.  This fact supports a proof that, in the chiral limit, excited state $0^-$ mesons do not couple to the axial-vector current; viz., $f_{\pi_n}\equiv 0$ $\forall n \geq 1$.  

We demonstrated that the leading-order (rainbow-ladder) term in the DSE truncation scheme, when consistently implemented, is necessary and sufficient to express the Abelian anomaly.  It can therefore be used to illustrate the anomaly's observable consequences.  We capitalised on this to show that even though excited state pseudoscalar mesons decouple from the axial-vector current in the chiral limit, they nevertheless couple to two photons.  (NB.\ The strength of this coupling is materially affected by the continuum contribution to the Abelian anomaly.)  Hence the Primakov process, as employed for example in \emph{PrimEx} at JLab \cite{primex}, may be used as a tool for their production and study.

A renormalisation-group-improved rainbow-ladder trun\-cation is guaranteed to express the one-loop renormalisation group properties of QCD.  We exploited this and thereby determined the leading power-law behaviour of the $\gamma^\ast \pi_n \gamma^\ast$ transition form factor.  When the current-quark mass is nonzero then, for all $n$, this form factor behaves as $(4\pi^2/3) (f_{\pi_n}/Q^2)$ at deep spacelike momenta.  For all but the Goldstone mode this leading order contribution vanishes in the chiral limit.  In that case, however, the form factor remains nonzero and the ultraviolet behaviour is $\simeq (4\pi^2/3) (-\langle \bar q q \rangle/Q^4)$.  Although only exposed starkly in the chiral limit for excited states, this subleading power-law contribution to the $\gamma^\ast \pi_n \gamma^\ast$ transition form factor is always present and in general its coefficient is not simply related to $f_{\pi_n}$.

As one might rationally expect, the properties of excited ($n\geq 1$) states are sensitive to the pointwise behaviour of what might be called the confinement potential between light-quarks.  We illustrated this by laying out the evolution of the charge radii of the $n=0,1$ pseudoscalar mesons.  As it is shielded by Goldstone's theorem, the ground state's radius can be insensitive to details of the long-range part of the interaction.  However, that is not true of $r_{\pi_1}$, the radius of the first excited state, which is orthogonal to the vacuum.  An increase in the length-scale that characterises the range of the confining potential reduces $r_{\pi_1}$.  This result states that increasing the confinement force compresses the excited state: indeed, it is possible to obtain $r_{\pi_1} < r_{\pi_0}$.  However, our current best estimate is $r_{\pi_1} \simeq 1.4\, r_{\pi_0}$.

A detailed exploration of the properties of collections of mesons on particular $J^P$ trajectories offers the hope of exposing features of the long-range part of the interaction between light-quarks.  In principle, this interaction can be quite different to that between heavy-quarks.  The pseudoscalar trajectory is of particular interest because its lowest mass entry is QCD's Goldstone mode. Chiral current conservation places constraints on some properties of every member of this trajectory, whose study may therefore provide information about the interplay between confinement and dynamical chiral symmetry breaking.

\bigskip

\centerline{Acknowledgments}\medskip
We acknowledge profitable interactions with S.\,J.~Brodsky, R.\,J.\ Holt and P.\,C.\ Tandy.
This work was supported by: Austrian Research Foundation \textit{FWF, 
Erwin-Schr\"odinger-Stipendium} no.\ J2233-N08; Department of Energy, 
Office of Nuclear Physics, contract nos.\ W-31-109-ENG-38 and DE-FG02-00ER41135; National Science Foundation contract no.\ INT-0129236; the \textit{A.\,v.\ 
Humboldt-Stiftung} via a \textit{F.\,W.\ Bessel Forschungspreis}; and benefited from the facilities of the ANL Computing Resource Center and the NSF Terascale Computing System at the Pittsburgh Supercomputing Center.
%
%

\nopagebreak

\appendix*
\section{Ultraviolet behaviour of transition form factor}
\label{app:a}
We observed in Sec.\,\ref{exactUV} that since we are concerned with $J^{PC} = 0^{-+}$ states it follows that 
\begin{eqnarray}
\nonumber
\lefteqn{ T^{\pi_n^0}_{\mu\nu}(k_1,k_2) = \frac{4 \pi \alpha}{3}\, i\varepsilon_{\mu\nu\rho\sigma}\,}\\
&=&  
 \left[K_\rho {\cal I}_\sigma(K,P) - K_\alpha {\cal J}_{\rho\sigma\alpha}(K,P)\right],
\label{TanswerA}
\end{eqnarray}
where Eq.\,(\ref{TpinggNa}) yields
\begin{eqnarray}
\nonumber
\lefteqn{{\cal I}_\sigma(K,P) }\\
\nonumber
&= & {\rm tr} \, Z_2 \int_\ell^{M} \!\! \sfrac{1}{2} \tau^3\, \gamma_5 \gamma_\sigma \, \chi_{\pi_n^0}(\ell;P) \,(\ell^2 + K^2) \, \Delta(\ell,K),  \\ \label{IresA}
\end{eqnarray}
with $ \Delta(l,K) = 1/[(\ell^2+K^2)^2 - 4 (\ell\cdot K)^2]$.  In this appendix we work always on the ultraviolet domain where $K\cdot P=0$ and $K^2=Q^2$.  

The leading contribution to $T^{\pi_n^0}_{\mu\nu}(k_1,k_2)$ is obtained from Eq.\,(\ref{IresA}).  That is apparent because 
\begin{equation}
(\ell^2 + K^2) \, \Delta(\ell,K) = \frac{1}{Q^2} + {\rm O}\left(\frac{1}{Q^4}\right),
\end{equation}
and this result inserted into Eq.\,(\ref{IresA}), with Eq.\,(\ref{fpin}) used to identify the residue, yields 
\begin{equation}
{\cal I}_\sigma(K,P) = P_\sigma \left\{\frac{f_{\pi_n}}{Q^2} + {\rm O}\left(\frac{1}{Q^4} \right)\right\}.
\end{equation}
Equation~(\ref{TpiuvQCD}) follows immediately.

Herein we also want the subleading contribution.  Consider Eq.\,(\ref{IresA}) with the explicit $K^2$ factor, which we have already used, removed from the numerator:
\begin{equation}
\tilde{\cal I}_\sigma(K,P) = 
 {\rm tr} \, Z_2 \int_\ell^{M} \!\! \sfrac{1}{2} \tau^3\, \gamma_5 \gamma_\sigma \, \chi_{\pi_n^0}(\ell;P) \,\ell^2  \, \Delta(\ell,K) \,.
\end{equation}
This term's contribution to Eq.\,(\ref{TanswerA}) can be written  
\begin{equation}
\tilde{\cal I}_\sigma(K,P)  =  P_\sigma \, {\cal F}_{\cal I}^{(2)}(P^2,K^2) \,,
\end{equation}
with 
\begin{eqnarray}
\nonumber
\lefteqn{ {\cal F}_{\cal I}^{(2)}(P^2,K^2) \, P^2 }\\
=& & \!\!\!\! {\rm tr} \, Z_2 \int_\ell^{M} \!\! \sfrac{1}{2} \tau^3\, \gamma_5 \gamma\cdot P \, \chi_{\pi_n^0}(\ell;P) \,\ell^2  \, \Delta(\ell,K) \,. \label{FINLO}
\end{eqnarray}

Inspection and consideration of Eqs.\,(\ref{definechi}), (\ref{sinvp}), (\ref{genpibsa}) reveals that one may write 
\begin{eqnarray}
{\rm tr}_{\rm D}[ \gamma_5 \gamma\cdot P \, \chi_{\pi_n^0}(\ell;P)] = 4\, P^2\,{\cal X}_{P^2}(\ell^2,(\ell\cdot P)^2),
\end{eqnarray}
where $P^2=-m_{\pi_n}^2$; i.e., $P^2$ is an eigenvalue, not a variable, and the result is a function of $(\ell\cdot P)^2$ because $J^{PC}=0^{-+}$.  (NB.\ For the following argument it is not necessary to make explicit the colour and flavour structure.)  

It is further apparent from Eqs.\,(\ref{definechi}), (\ref{sinvp}), (\ref{genpibsa}) that when chiral symmetry is dynamically broken
\begin{equation}
0<{\cal X}_{P^2}(0,0)<\infty\,;
\end{equation}
and, moreover, that ${\cal X}_{P^2}(\ell^2,(\ell\cdot P)^2)$ is a smooth function of its arguments so that it may be written
\begin{equation}
{\cal X}_{P^2}(\ell^2,\ell\cdot P) = \sum_{i=0}^\infty \, {\cal X}_{P^2}^i(\ell^2) \,  (\ell\cdot P)^{2 i}\,.
\end{equation}
In addition, 
\begin{equation}
\label{Xuv}
{\cal X}_{P^2}^i(\ell^2) \stackrel{\ell^2\gg\omega_{\pi_n}^2}{\sim} \left(\frac{1}{\ell^2}\right)^{(3+i)},
\end{equation}
up to $\ln^{\tilde\gamma}(\ell^2/\tilde\omega_{\pi_n}^2)$-corrections, where $\tilde\omega_{\pi_n}^2$ is a mass-scale that characterises the momentum-space width of the pseudoscalar meson's Bethe-Salpeter wave function and $\tilde\gamma$ is this wave function's anomalous dimension.  

One can now return to Eq.\,(\ref{FINLO}) and use the information provided to determine the dominant contribution 
\begin{equation}
{\cal F}_{\cal I}^{(2)}(P^2,K^2) = {\rm tr} \, Z_2 \int_\ell^{M} \!\! \sfrac{1}{2} \tau^3\, {\cal X}^0_{P^2}(\ell^2) \,\ell^2  \, \Delta(\ell,K)\,.
\end{equation}
The angular integration is straightforward:
\begin{equation}
\label{anglekernel}
\int d^4\Omega_\ell\, \Delta(\ell,K) = \frac{\pi^2}{K^2 \ell^2} \frac{K^2 + \ell^2 - |K^2 - \ell^2|}{K^2+\ell^2}\,,
\end{equation}
from which it follows that (recall $K^2=Q^2$)
\begin{eqnarray}
\nonumber
\lefteqn{
{\cal F}_{\cal I}^{(2)}(P^2,Q^2)}\\
\nonumber 
&& \stackrel{Q^2\to\infty}{=} \frac{1}{Q^2} \, {\rm tr} \, Z_2 \frac{1}{8\pi^2} \int_0^{Q^2}\!\! dy\, \sfrac{1}{2} \tau^3\, {\cal X}^0_{P^2}(y) \frac{y^2}{Q^2+y}\\
\label{lnUV}
&& \stackrel{Q^2\to\infty}{=} F^{(2)}_{\cal I}(P^2) \, \frac{\ln(Q^2/\tilde\omega_{\pi_n}^2)}{Q^4}\,,
\end{eqnarray}
where we have used Eq.\,(\ref{Xuv}) and neglected the anomalous dimension.  (NB.\ An inspection of Eq.\,(\ref{anglekernel}) will reveal that the contribution to the integral from $\ell^2>Q^2$ is finite and, as $Q^2\to\infty$, suppressed with respect to the term we have exposed.)  

The true exponent and width in the logarithmic modification to the power-law behaviour will be affected by, e.g., dressing of the quark-gluon vertex; i.e., corrections to Eq.\,(\ref{expand}), and diagrams beyond the rainbow-ladder truncation.  Herein we are satisfied merely to establish that the subleading power-law behaviour is $1/Q^4$ and that, in general, a $\ln^{\gamma} Q^2$-modification may be present.  

The analysis presented in this appendix can be adapted to show that in general the leading contribution from $K_\alpha {\cal J}_{\rho\sigma\alpha}(K,P)$ in Eq.\,(\ref{Jres}) also exhibits behaviour of the type in Eq.\,(\ref{lnUV}).


\begin{thebibliography}{00}
\bibitem{fn:Eucl} We employ a Euclidean metric, with:  $\{\gamma_\mu,\gamma_\nu\} = 2\delta_{\mu\nu}$; $\gamma_\mu^\dagger = \gamma_\mu$; $ \gamma_5 = - \gamma_1 \gamma_2 \gamma_3 \gamma_4$; ${\rm tr}\, \gamma_5 \gamma_\mu \gamma_\nu \gamma_\rho \gamma_\sigma = -4\, \varepsilon_{\mu\nu\rho\sigma}$;  and $a \cdot b = \sum_{i=1}^4 a_i b_i$.  For a timelike vector $P_\mu$, $P^2<0$.
\bibitem{fn:landau} We use Landau gauge in all calculations because it is a fixed-point of the renormalisation group.
\end{thebibliography}
\end{document}